\documentclass[prb,aps,twocolumn,showpacs,floats]{revtex4}
\usepackage{epsfig,bm,epsf,graphics}
\begin{document}
\draft
\title{FRUSTRATION EFFECTS IN ANTIFERROMAGNETIC FCC HEISENBERG FILMS}
\author{V. Thanh Ngo$^{a,b}$ and H. T. Diep\footnote{ Corresponding author, E-mail:diep@u-cergy.fr }}
\address{Laboratoire de Physique Th\'eorique et Mod\'elisation,
CNRS-Universit\'e de Cergy-Pontoise, UMR 8089\\
2, Avenue Adolphe Chauvin, 95302 Cergy-Pontoise Cedex, France\\
$^a$ Institute of Physics, P.O. Box 429,   Bo Ho, Hanoi 10000,
Vietnam\\
$^b$ Asia Pacific Center for Theoretical Physics, Hogil Kim
Memorial Building 5th floor, POSTECH, Hyoja-dong, Namgu, Pohang
790-784, Korea}

\begin{abstract}
We study the effects of frustration in an antiferromagnetic film
of FCC lattice with Heisenberg spin model including an Ising-like
anisotropy.  Monte Carlo (MC) simulations have been used to study
thermodynamic properties of the film.  We show that the presence
of the surface reduces the ground state (GS) degeneracy found in
the bulk.  The GS is shown to depend on the  surface in-plane
interaction $J_s$ with a critical value at which ordering of type
I coexists with ordering of type II. Near this value a reentrant
phase is found.  Various physical quantities such as layer
magnetizations and layer susceptibilities are shown and discussed.
The nature of the phase transition is also studied by histogram
technique. We have also used the Green's function (GF) method for
the quantum counterpart model.  The results at low-$T$ show
interesting effects of quantum fluctuations.  Results obtained by
the GF method at high $T$ are compared to those of MC simulations.
A good agreement is observed.
\end{abstract}
\maketitle
\section{Introduction}
Effects of the frustration in spin systems have been extensively
investigated during the last 30 years.  Frustrated spin systems are shown to
have unusual properties such as large ground state (GS)
degeneracy, additional GS symmetries, successive phase transitions
with complicated nature.  Frustrated systems still challenge
theoretical and experimental methods. For  recent reviews, the
reader is referred to Ref. \onlinecite{Diep2005}.

On the other hand, during the same period physics of surfaces and objects of nanometric
size have also attracted an immense interest.  This is
due to important applications in
industry.\cite{zangwill,bland-heinrich,Binder-surf,Diehl} In this
field, results from laboratory research are often immediately used
for industrial applications, without waiting for a full
theoretical understanding. An example is the so-called giant
magneto-resistance (GMR) used in data storage devices, magnetic sensors,
... \cite{Baibich,Grunberg,Fert,review}  In parallel to these 
experimental developments, much theoretical effort has also been devoted to
the search of physical mechanisms lying behind new properties found
in nanometric objects such as ultrathin films, ultrafine particles, quantum dots,
spintronic devices etc. This effort aimed not only at providing explanations for
experimental observations but also at predicting new effects for future 
experiments.\cite{ngo2004trilayer,ngo2007}

The aim of this paper is to investigate the effect of the presence
of a film surface in a system which is known to be very
frustrated, namely the FCC antiferromagnet. The bulk properties of this
material have been largely studied as we will show below.  In this paper,  we
would like to see in particular how the  frustration effects on the nature of the
phase transition in 3D are modified in  thin films and how the surface
conditions affect the magnetic phase diagram.  To carry out these
purposes, we shall use Monte Carlo (MC) simulations and the
Green's function (GF) method for qualitative comparison.

The paper is organized as follows. Section II is devoted to the
description of the model. We recall there properties of the 3D
counterpart model in order to better appreciate properties of thin
films obtained in this paper. A determination of its GS properties
is also given.  In section III, we show our results obtained by MC
simulations as functions of temperature $T$. The surface exchange
interaction $J_s$ is made to vary. A phase diagram in the space
$(T,J_s)$ is shown and discussed.  In general, the surface 
transition is found to be distinct from the transition of interior layers. 
An interesting reentrant region is observed in the phase diagram.  
We also show in this section the
results on the critical exponents obtained by MC multihistogram technique.
A detailed discussion on the
nature of the phase transition is given.  Section IV is devoted to
a study of the quantum version of the same model by the use of the
GF method. We find interesting effects of quantum fluctuations at low 
$T$. The phase diagram $(T,J_s)$ is
established and compared to that obtained by MC simulations for
the classical model. Concluding remarks are given in section V.

\section{Model and Classical Ground State Analysis}

It is known that the antiferromagnetic (AF) interaction between
nearest-neighbor (NN) spins on the FCC lattice causes  a very
strong frustration.  This is due to the fact that the FCC lattice
is composed of corner-sharing tetrahedra each of which has four
equilateral triangles.  It is well-known\cite{Diep2005} that it is
impossible to fully satisfy simultaneously the three AF bond
interactions on each triangle.

The analytical determination  of the GS of systems of classical
spins with competing interactions is a fascinating subject. For a
recent review, the reader is referred to Ref. \onlinecite{Kaplan}.
For the bulk FCC antiferromagnet, the Heisenberg spins on a
tetrahedron form a configuration characterized by two arbitrary
angles.\cite{Oguchi} The ground state (GS) degeneracy is therefore
infinite. This is also found in fully frustrated simple cubic
lattice with classical Heisenberg spins:\cite{Lallemand1985ff} the
GS is also characterized by two random continuous parameters.  To
give an idea about the GS of the bulk FCC
antiferromagnet,\cite{Oguchi} let us imagine two planes, $xz$ and
$\psi$,  where $\psi$ intersects with the $xz$ plane along the $z$
axis and makes an angle $\phi$ with the $x$ axis. Two of the four
spins make an angle $\theta$ in the $xz$ plane symmetric with
respect to the $z$ axis. The other two spins make also the same
angle, symmetric with respect to the $z$ axis, but in the plane
$\psi$. It has been shown\cite{Oguchi} that the two angles
$\theta$ and $\phi$ are arbitrary between 0 and $\pi$. Note that
when $\theta=0$ the spin configuration is collinear with two spins
along the $+z$ axis and the other two along the $-z$ one. The
phase transition of the bulk frustrated FCC Heisenberg
antiferromagnet has been studied.\cite{Diep1989fcc} In particular,
the transition is found to be of the first order as in the Ising
case.\cite{Phani,Polgreen,Styer} Other similar frustrated
antiferromagnets such as the HCP antiferromagnet show the same
behavior.\cite{Diep1992hcp}

Let us consider a film of FCC lattice structure with (001)
surfaces. To avoid the absence of long-range order of isotropic
non Ising spin model at finite temperature ($T$) when the film
thickness is very small, i.e. quasi two-dimensional
system\cite{Mermin}, we add in the Hamiltonian an Ising-like
uniaxial anisotropy term. The Hamiltonian is given by
\begin{equation}
\mathcal H=-\sum_{\left<i,j\right>}J_{i,j}\mathbf S_i\cdot\mathbf
S_j -\sum_{<i>} D_{i}(S^z_i)^2  \label{eqn:hamil1}
\end{equation}
where $\mathbf S_i$ is the Heisenberg spin at the lattice site
$i$, $\sum_{\left<i,j\right>}$ indicates the sum over the NN spin
pairs  $\mathbf S_i$ and $\mathbf S_j$.  

In the following, the interaction between two NN surface spins is denoted by $J_s$, while
all other interactions are supposed to be antiferromagnetic and all equal to $J=-1$ for
simplicity. 

\begin{figure}
\centerline{\epsfig{file=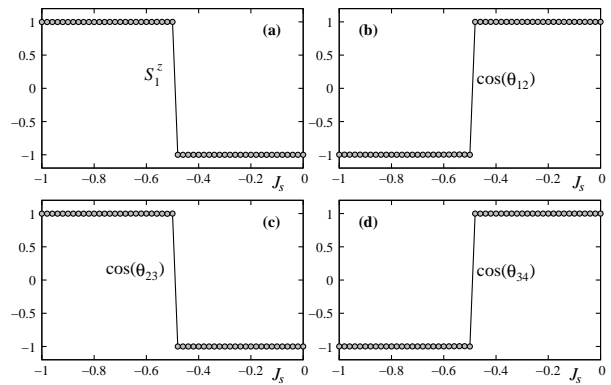,width=3.2in}} \caption{A
ground state configuration of single plaquette (a) $S^z_1$ is $S^z$
of sublattice 1, (b) $\cos\theta_{12}$, (c) $\cos\theta_{23}$, (d)
$\cos\theta_{34}$.  $\cos \theta_{ij}$   is cosine of angle between the two spins of  
sublattices $i$ and $j$.} \label{fig:gscos}
\end{figure}

We first determine the GS configuration by
using the steepest descent method : starting from a random spin
configuration, we calculate the magnetic local field at each site
and align the spin of the site in its local field. In doing so for
all spins and repeat until the convergence is reached, we obtain easily
the GS configuration without metastable states.  The result is shown in Fig.
\ref{fig:gscos}


\begin{figure}
\centerline{\epsfig{file=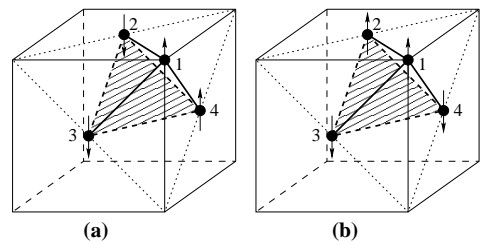,width=2.5in}} \caption{The
ground state spin configuration of the FCC cell at the film
surface: a) ordering of type I for $J_s<-0.5$ b) ordering of type
II for $J_s>-0.5$.} \label{fig:gsstruct}
\end{figure}

We observe that there is a critical value $J^c_s =-0.5$. For $J_s < J^c_s$, the
spins in each $yz$ plane are parallel while spins in adjacent $yz$
planes are antiparallel (Fig. \ref{fig:gsstruct}a). This ordering
will be called hereafter "ordering of type I": in the $x$ direction the 
ferromagnetic planes are antiferromagnetically coupled as shown in this
figure. Of course, there is a degenerate configuration where the ferromagnetic planes are
antiferromagnetically ordered in the $y$ direction.  Note that the surface layer
has an AF ordering for both configurations.  The degeneracy of  type
I is therefore 4 including the reversal of all spins.

For $J_s
> J^c_s$, the spins in each $xy$ plane is
ferromagnetic. The adjacent $xy$ planes have an AF ordering in the $z$
direction perpendicular to the film surface.   This
will be called hereafter "ordering of type II".  Note that the
surface layer is then ferromagnetic (Fig. \ref{fig:gsstruct}b). The
degeneracy of  type II is 2 due to the reversal of all spins.

Without using a general method,\cite{Kaplan,Oguchi} let us
calculate analytically the GS configuration in a simple manner for
the present model.

Consider a tetrahedron with the spins numbered as in Fig.
\ref{fig:gsstruct}: $S_1$, $S_2$, $S_3$ and $S_4$ are the spins in
the surface FCC cell (first cell). The interaction between $S_1$
and $S_2$ is set to be equal to $J_s$ $(-1 \leq J_s \leq 0)$ and all others
are taken to be equal to $J$ ($<0$), and all $D_i=D$ for simplicity. The
Hamiltonian for the cell is written as
\begin{eqnarray}
H_p &=& -\frac{1}{2}\Big\{ 8 J_s (\mathbf S_1\cdot \mathbf S_2) + 8 J (\mathbf S_3\cdot \mathbf S_4)\nonumber\\
&+& 6 J \left[\mathbf S_1\cdot\mathbf S_3 + \mathbf S_1\cdot\mathbf S_4 +\mathbf S_2\cdot\mathbf S_3 + \mathbf S_2\cdot\mathbf S_4
\right]\nonumber\\
&+& 2D \left[ (S^z_1)^2+(S^z_2)^2+(S^z_3)^2+(S^z_4)^2\right]\Big\}. \label{eqn:Hamilplaq}
\end{eqnarray}
Let us decompose each spin into two components: an $xy$ component,
which is a vector, and a $z$ component $\mathbf S_i=(\mathbf
S_i^{\parallel}, S_i^z)$. The numerical results shown above indicate that
the spins have only $z$ component.  Taking advantage of this, we suppose that the $xy$ vector
components of the spins are all equal to zero. 
The angles $\theta_i$ of $\mathbf S_i$ with the $z$ axis are then
$$
\left\{%
\begin{array}{ll}
\theta_1=0, &\theta_3=\pi\\
\theta_2=\beta_1,& \theta_4=\beta_2.
\end{array}
\right.
$$

 The total energy
of the cell (\ref{eqn:Hamilplaq}), with $S_i =
\frac{1}{2}$, can be rewritten as
\begin{eqnarray}
H_p&=&-\frac{D}{2}+\frac{3J}{4}+\left(\frac{3J}{4}-J_s-\frac{D}{4}\cos\beta_1\right)\cos\beta_1\nonumber\\
&+&\frac{1}{4}\left(J-D\cos\beta_2\right)\cos\beta_2-\frac{3J}{4}\cos\beta_1\cos\beta_2. \label{eqn:totEplaq}
\end{eqnarray}
By a variational method,  the minimum of the cell energy
corresponds to
\begin{eqnarray}
\label{eqn:DerivE1}
\frac{\partial H_p}{\partial\beta_1}&=&J_s\sin\beta_1+\frac{D}{2}\cos\beta_1\sin\beta_1\nonumber\\
&-&\frac{3J}{4}\sin\beta_1+\frac{3J}{4}\cos\beta_2\sin\beta_1 = 0,\\
\label{eqn:DerivE2}
\frac{\partial H_p}{\partial\beta_2}&=&\left[\frac{3J}{4}\cos\beta_1-\frac{J}{4} +\frac{D}{2}\cos\beta_2\right]\sin\beta_2= 0.
\end{eqnarray}
The solutions of Eq. (\ref{eqn:DerivE1}) and Eq.
(\ref{eqn:DerivE2}) corresponding to the minimal energy are
\begin{equation}
\left\{
\begin{array}{ll}
\cos\beta_1 = -\cos\beta_2=-1&\mathrm {for}\ |J_s| > 0.5,\\
\cos\beta_1 = -\cos\beta_2=1 &\mathrm {for}\ |J_s| <  0.5.
\end{array}
\right.
\label{eqn:GSsolu}
\end{equation}
Note that these solutions do not depend on $D$.   The GS energy
per spin is
\begin{equation}
\left\{
\begin{array}{ll}
H_p=-D + J + J_s &\mathrm {for}\ |J_s| > 0.5,\\
H_p=-D + 2J-J_s  &\mathrm {for}\ |J_s| <  0.5.
\end{array}
\right.
\label{eqn:GSEsolu}
\end{equation}


We see that the solution (\ref{eqn:GSsolu}) agrees with the numerical result.

\section{Monte Carlo Results}
In this paragraph, we show the results obtained by MC simulations
with the Hamiltonian (\ref{eqn:hamil1}). The spins are the
classical Heisenberg model of magnitude $S=1$.

 The film size is $L\times L \times N_z$ where
$N_z$ is the number of FCC cells along the $z$ direction (film
thickness). Note that each cell has two atomic planes.  We use
here $L=12,18,24, 30,36$ and $N_z=$ 4. Periodic boundary
conditions are used in the $xy$ planes. The equilibrating time is
about $10^6$ MC steps per spin and the averaging time is $2\times
10^6$ MC steps per spin. $|J|=1$ is taken as unit of energy in the
following.


Before showing the results let us adopt the following notations.
The sublattice 1 of the first cell belongs to the surface layer,
while the sublattice 3 of the first cell belongs to the second
layer.  The sublattices 1 and 3 of the second cell belong,
respectively, to the third and fourth layers. In our simulations,
we used four cells, $N_z=4$, i.e. 8 atomic layers.  The symmetry
of the two film surfaces imposes the equivalence of the first and
fourth cells and that of the second and third cells.  It suffices
then to show results of the first two cells, i. e. four first
layers.  In addition, in each atomic layer the two sublattices are
equivalent by symmetry.  Therefore, we choose  to show in the
following the results of the sublattices 1 and 3 of the first two
cells, i.e. results of the first four layers.

\begin{figure}
\centerline{\epsfig{file=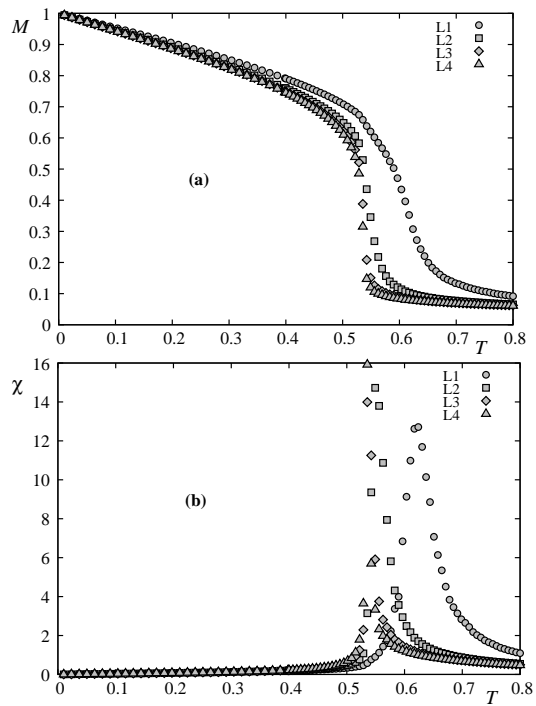,width=2.8in}}
\caption{Magnetizations and susceptibilities of sublattices 1 and
3 first two cells vs temperature for $J_s = -1.0$ with $L=24$ and
$D = 0.1$. $L_j$ denotes the sublattice magnetization of layer
$j$.} \label{fig:N24D01J10}
\end{figure}

Let us show in Fig. \ref{fig:N24D01J10}  the magnetizations and
the susceptibilities of sublattices 1 and 3 of the first two
cells, in the case where $J_s=-1$.

\begin{figure} [h!]
\centerline{\epsfig{file=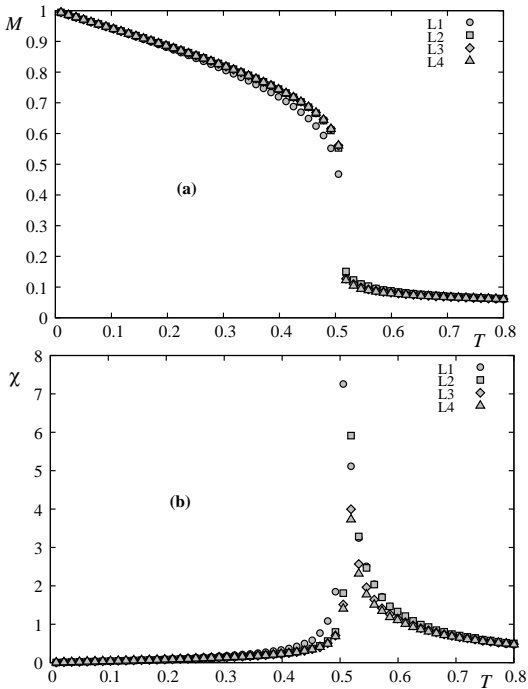,width=2.8in}}
\caption{Magnetizations and susceptibilities of sublattices 1 and
3 of first two cells vs temperature for $J_s = -0.8$ with $L=24$
and $D = 0.1$. $L_j$ denotes the sublattice magnetization of layer
$j$.} \label{fig:N24D01J08}
\end{figure}

\begin{figure} 
\centerline{\epsfig{file=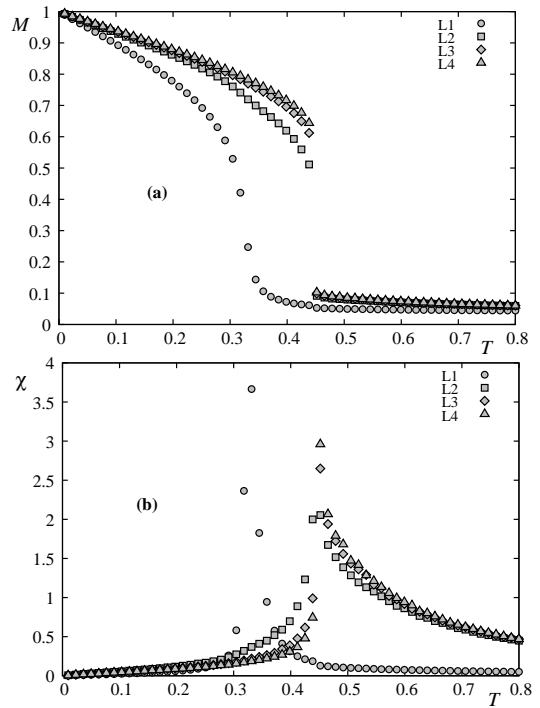,width=2.8in}}
\caption{Magnetizations and susceptibilities of first two cells vs
temperature for $J_s = -0.5$ with $L=24$ and $D = 0.1$. $L_j$
denotes the sublattice magnetization of layer $j$. The
susceptibility of sublattice 1 of the first cell is divided by a
factor $5$ for presentation convenience.} \label{fig:N24D01J05}
\end{figure}

It is interesting to note that the surface layer has largest
magnetization followed by that of the second layer, while the
third and fourth layers have smaller magnetizations.  This is not
the case for non frustrated films where the surface magnetization
is always smaller than the interior ones because of the effects of
low-lying energy surface-localized magnon
modes.\cite{diep79,diep81}  One explanation can be advanced: due
to the lack of neighbors surface spins suffer fluctuations due to
the frustration less than the interior spins so they maintain
their ordering up to a higher temperature.  Let us decrease the
$J_s$ strength. The surface spins then have smaller local field,
so thermal fluctuations will reduce their ordering to a lower
temperature. Figures \ref{fig:N24D01J08} and \ref{fig:N24D01J05}
show respectively the cases where $J_s=-0.8$ and -0.5.  Near
$J_s=-0.8$ the crossover takes place: the surface magnetization
becomes smaller than the interior ones for $J_s
>-0.8$.  Note that the magnetizations of second, third and
fourth layers undergo a discontinuity at the transition
temperature for $J_s=-0.8$ and -0.5.  This suggests that the phase
transitions for interior layers are of first order as it has been
found for bulk FCC antiferromagnet.\cite{Diep1989fcc} This should
be checked in the future.

\begin{figure}
\centerline{\epsfig{file=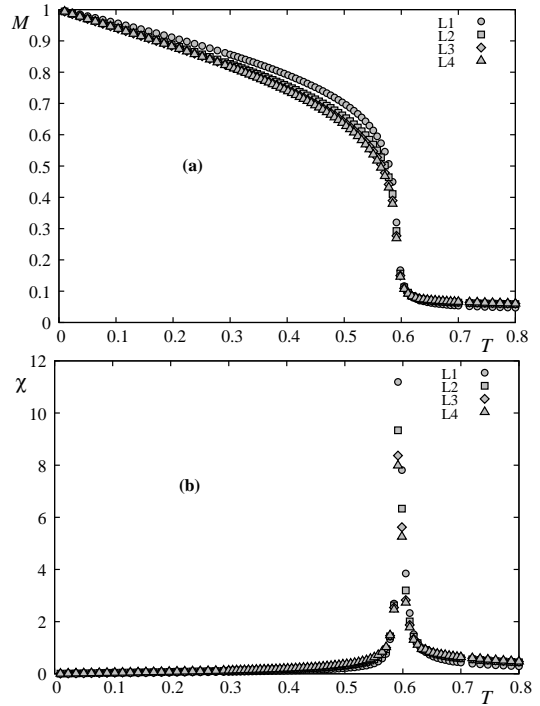,width=2.8in}}
\caption{Magnetization and susceptibility of first two cells vs
temperature for $J_s = -0.1$ with $L=24$ and $D = 0.1$. $L_j$
denotes the sublattice magnetization of layer $j$.}
\label{fig:N24D01J01}
\end{figure}

For weak $|J_s|$, there is only one transition for all layers. An
example is shown in Fig. \ref{fig:N24D01J01} for $J_s=-0.1$. Note
that the first-order character disappears as there is no
discontinuity of layer magnetizations at the transition
temperature.

\begin{figure}
\centerline{\epsfig{file=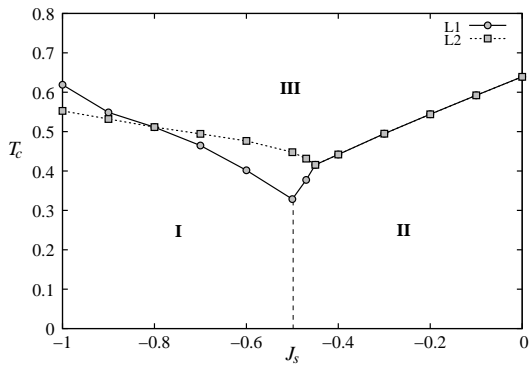,width=2.8in}}
\caption{Critical temperature vs $J_s$ with $L=24$ and $D = 0.1$.
$L_j$ denotes data points for the sublattice magnetization of
layer $j$.  I and II denote ordering of type I and II defined in
in Fig. \ref{fig:gsstruct}.  III is paramagnetic phase.  The
discontinued vertical line is a first-order line. See text for
comments. } \label{fig:PDJ24D01}
\end{figure}

In the region  $-0.5 <J_s< -0.45$, there is an interesting
reentrant phenomenon.  To facilitate the description of this
phenomenon, let us show the phase diagram in the space ($J_s,T_c$)
in Fig. \ref{fig:PDJ24D01}. In the region $-0.5 <J_s< -0.45$, the
GS is of type II as seen above. According to the phase diagram, we
see that when the temperature increases from zero, the system goes
through the phase of type II, undergoes a transition to enter the
phase of type I before making a second transition to the
paramagnetic phase at high temperature. This kind of behavior is
termed as reentrant phenomenon which has been found by exact
solutions in a number of very  frustrated
systems.\cite{diep91honeyc,diep91Kago} For a complete review on
these exactly solved systems, the reader is referred to the
chapter by Diep and Giacomini\cite{Diep-Giacomini} in Ref.
\onlinecite{Diep2005}.  We note here that the reentrance is often
found near the frontier where two phases coexist in the
GS.\cite{Diep2005} This is the case at $J_s=J_s^c=-0.5$.

\begin{figure}
\centerline{\epsfig{file=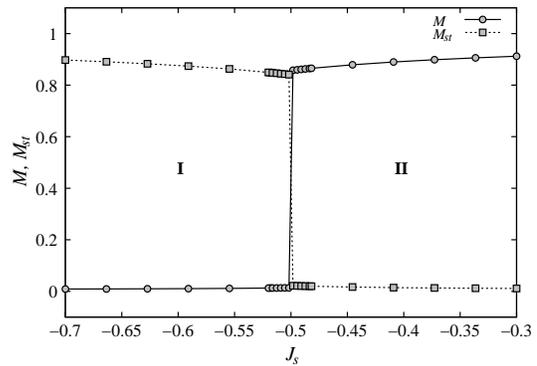,width=2.8in}} \caption{The
magnetization $M$ and the staggered magnetization $M_{st}$  of
first layer versus $J_s$ are shown, at $T=0.15$, with $L=24$ and
$D = 0.1$. I and II denote ordering of type I and II defined  in
Fig. \ref{fig:gsstruct}. III is paramagnetic phase. See text for
comments. } \label{fig:N24Order}
\end{figure}

The discontinued vertical line at $J_s=-0.5$ is a first-order line
separating phases I and II. The coexistence of these two phases
which do not have the same symmetry explains the first-order
character of this line. To show it explicitly, we have calculated
at $T=0.15$ the magnetization $M$ and the staggered magnetization
$M_{st}$ of the first layer with varying $J_s$ across -O.5.  From
the GS configurations shown in Fig. \ref{fig:gsstruct}, $M$ should
be zero in phase I and finite in phase II, and vice-versa for
$M_{st}$. This is observed at $T=0.15$ as shown in Fig.
\ref{fig:N24Order}. The large discontinuity of $M$ and $M_{st}$ at
$J_s=-0.5$ shows a very strong first-order character across the
vertical line in Fig. \ref{fig:PDJ24D01}.

Let us discuss on finite-size effects in the transitions observed
in Fig. \ref{fig:N24D01J10} to Fig. \ref{fig:N24D01J01}.  This is
an important question because it is known that some apparent
transitions are  artifacts of small system sizes.  To confirm the
observed transitions, we have made a study of finite-size effects
on the layer susceptibilities by using the accurate MC multi
histogram technique.\cite{Ferrenberg1,Ferrenberg2,Ferrenberg3}

At this point, let us recall that bulk Ising frustrated systems,
unlike unfrustrated counterparts,  have different transition
natures:  the antiferromagnetic FCC and HCP Ising lattices have
strong first-order transition,\cite{Phani,Polgreen,Styer} while
the stacked antiferromagnetic triangular lattice has a
controversial nature (see references in Ref. \onlinecite{Plumer}).
The model studied here is the frustrated FCC film where surface
effects can modify the strong first-order observed in its bulk
counterpart.

\begin{figure}
\centerline{\epsfig{file=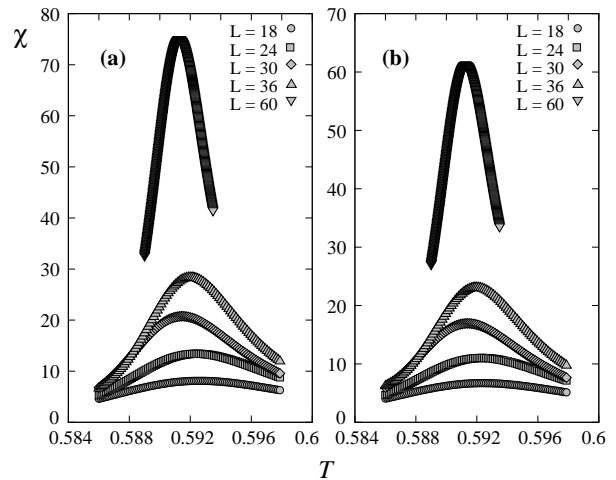,width=3.2in}}
\caption{Susceptibilities of layer 1 (left) and 2 (right) are
shown for various sizes $L$ as a function of temperature for
$J_s=-0.1$ and $D = 0.1$.} \label{fig:SUSJ01L12}
\end{figure}

\begin{figure}
\centerline{\epsfig{file=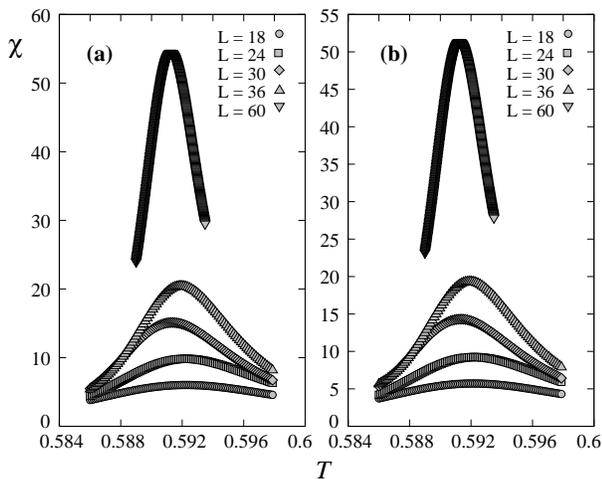,width=3.2in}}
\caption{Susceptibilities of layer 3 (left) and 4 (right) are
shown for various sizes $L$ as a function of temperature for
$J_s=-0.1$ and $D = 0.1$.} \label{fig:SUSJ01L34}
\end{figure}

Our results show that transitions at $J_s=-1$ and $J_s=-0.1$ are
real second-order transitions obeying some scaling law. Figure
\ref{fig:SUSJ01L12} shows the size effects on the maximum of the
susceptibilities of the first and second layers for $J_s=-0.1$,
while Fig. \ref{fig:SUSJ01L34} shows that of the third and fourth
layers.  As seen, the maximum of the susceptibilities
$\chi^{\max}$ increases with increasing $L$.

\begin{figure}
\centerline{\epsfig{file=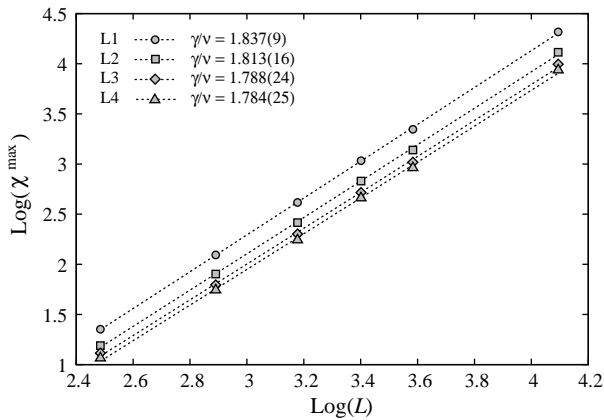,width=3.2in}} \caption{
Maximum sublattice susceptibility  $\chi^{\max}$ versus $L$ in the
$\ln-\ln$ scale, for $J_s=-0.1$ and $D = 0.1$. $L_j$ denotes the
sublattice magnetization of layer $j$. The slopes of these lines
give the ratios of exponents $\gamma/\nu$.} \label{fig:GAMJ01}
\end{figure}

Using the scaling law $\chi^{\max} \propto L^{\gamma/\nu}$, we
plot $\ln \chi^{\max}$  versus $\ln L$ in Fig. \ref{fig:GAMJ01}.
The ratio of the critical exponents $\gamma/\nu$ is obtained by
the slope of the straight line connecting the data points of each
layer.

\begin{figure}
\centerline{\epsfig{file=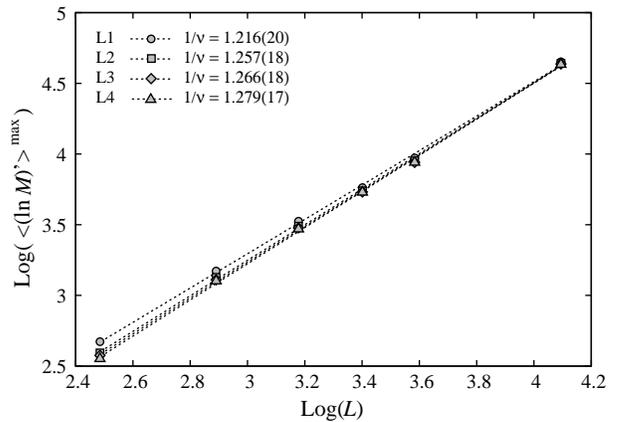,width=3.2in}} \caption{ The
maximum value of $\left<(\ln
M)'\right>=\left<E\right>-\left<ME\right>/\left<M\right>$ versus
$L$ in the $\ln-\ln$ scale for $J_s=-0.1$, where $M$ is the
sublattice order parameter. The slope of each line gives $1/\nu$.
$L_j$ denotes the sublattice magnetization of layer $j$.}
\label{fig:NUJ01}
\end{figure}

Within errors the third and fourth layers have the same value of
$\gamma/\nu$ which is neither 2D nor 3D Ising universality
classes, 1.75 and 2, respectively. The same for the the values of
the first and second layers. The  exponent $\nu$ can be obtained
as follows. We calculate as a function of $T$ the magnetization
derivative with respect to $\beta=(k_BT)^{-1}$: $V_1=\left<(\ln
M)'\right>=\left<E\right>-\left<ME\right>/\left<M\right>$ where
$E$ is the system energy and $M$ the sublattice order parameter.
We identify the maximum of $V_1$ for each size $L$. From the
finite-size scaling we know that $V_1^{\max}$ is proportional to
$L^{1/\nu}$.\cite{Ferrenberg3} We plot in Fig. \ref{fig:NUJ01}
$\ln V_1^{\max}$ as a function of $\ln L$ for $J_s=-0.1$. The
slope of each line gives $1/\nu$. For the case $J_s=-0.1$, we
obtain $\nu=0.822\pm 0.020,0.795\pm 0.020,0.790\pm 0.020, 0.782\pm
0.020$ for the first, second, third and fourth layers. These
values are far from the 2D value ($\nu=1$).  We deduce
$\gamma=1.510\pm 0.010,1.442\pm 0.015,1.412\pm 0.025,1.395\pm
0.025$. The values of $\nu$ and $\gamma$ are decreased when one
goes from the surface to the interior of the film.

\begin{figure}
\centerline{\epsfig{file=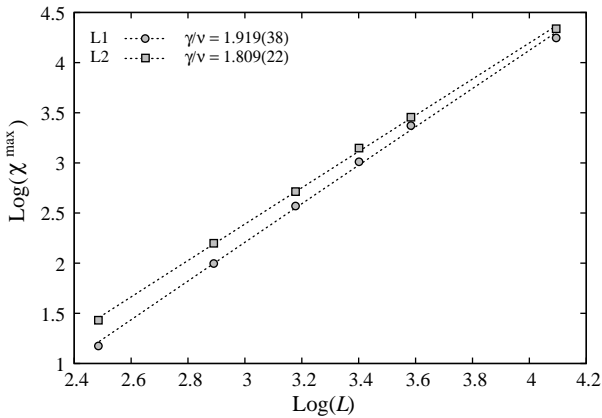,width=3.2in}} \caption{Maximum
sublattice susceptibility  $\chi^{\max}$ versus $L$ in the
$\ln-\ln$ scale, for $J_s=-1$ and $D = 0.1$. $L_j$ denotes the
sublattice magnetization of layer $j$. The slopes of these lines
give the ratios of exponents $\gamma/\nu$.} \label{fig:GAMJ10}
\end{figure}

\begin{figure}
\centerline{\epsfig{file=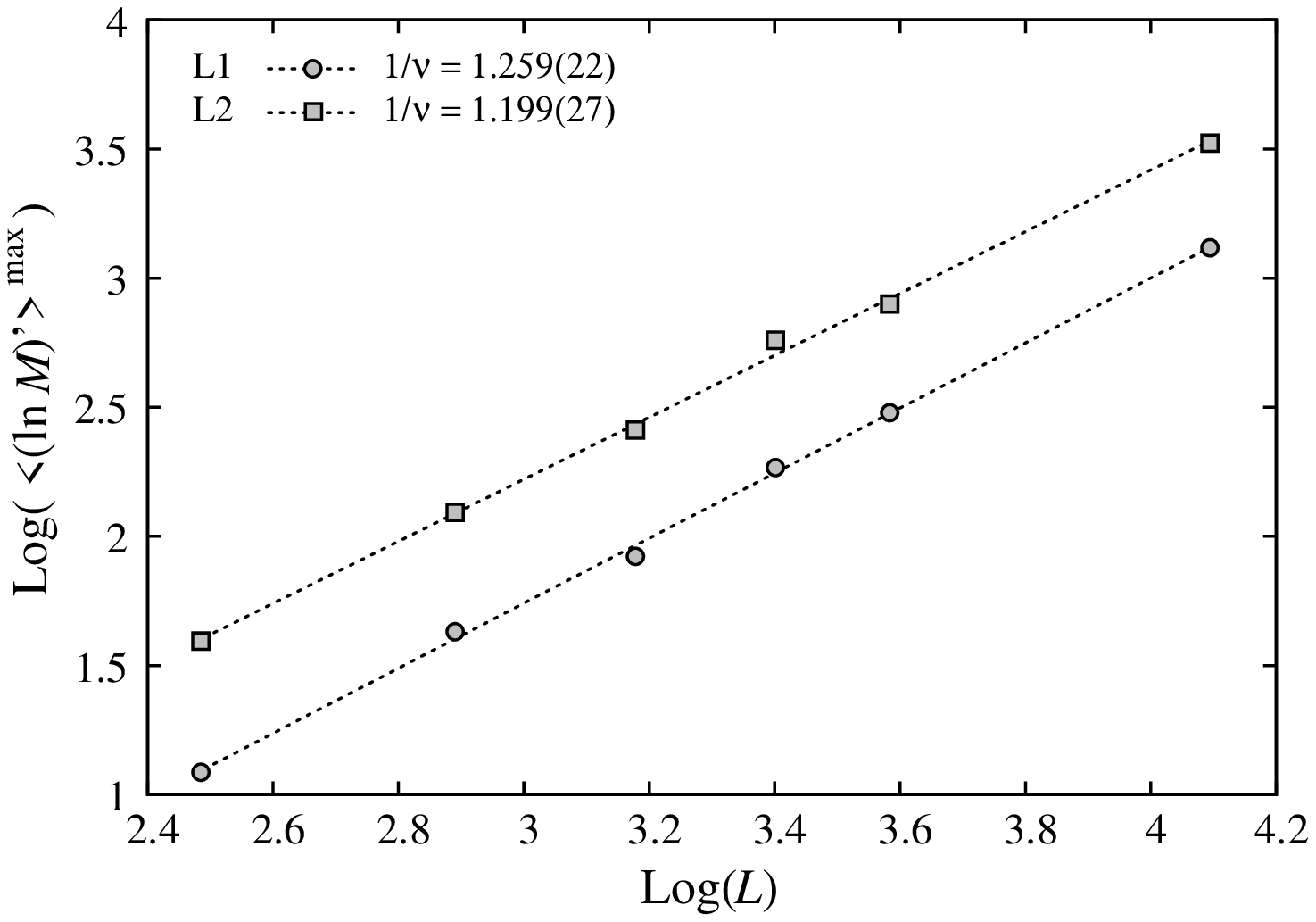,width=3.2in}} \caption{The
maximum value of $\left<(\ln
M)'\right>=\left<E\right>-\left<ME\right>/\left<M\right>$ versus
$L$ in the $\ln-\ln$ scale for $J_s=-1$, where $M$ is the
sublattice order parameter. The slope of each line gives $1/\nu$.
$L_kS_j$ denotes one sublattice magnetization of layer $j$.}
\label{fig:NUJ10}
\end{figure}

We show in Fig. \ref{fig:GAMJ10} and Fig. \ref{fig:NUJ10} the
maximum of sublattice magnetizations and their derivatives for the
first two layers in the case of $J_s=-1$.  We find $\nu_1=0.794\pm
0.022$, $\nu_2=0.834\pm 0.027$, $\gamma_1=1.524\pm 0.0.040$, and
$\gamma_2=1.509\pm 0.022$.

\vspace{10cm}

Let us discuss on the values of the critical exponents obtained
above. These values do not correspond neither to 2D nor 3D Ising
models ($\gamma_{2D}=1.75$, $\nu_{2D}=1$, $\gamma_{3D}=1.241$,
$\nu_{3D}=0.63$).  There are multiple reasons for those
deviations. Apart from numerical precisions and the modest sizes
we used, there may be deep physical origins.

A first question which naturally arises is the effect of the
frustration. The 3D version of this model, as said above, has a
first-order transition, with a very strong character for the Ising
case\cite{Phani,Polgreen,Styer} and somewhat less strong for the
continuous spin models.\cite{Diep1989fcc}  It has been shown that
at finite temperature, the phenomenon called "order by disorder"
occurs leading to a reduction of degeneracy: only collinear
configurations survive by an entropy
effect.\cite{Diep1989fcc,Villain,Henley}  The infinite degeneracy
is reduced to 6, i. e. the number of ways to put two AF spin pairs
on a tetrahedron.  The model is equivalent to 6-state Potts model.
 The first-order transition observed in the 3D case is in agreement
with the Potts criterion according to which the transition in
$q$-state Potts model is of first-order in 3D for $q\ge 3$.

In the case of a film with finite thickness studied here, it
appears that the first-order character is lost.

A first possible cause is from the degeneracy. According to the
results shown in the previous section, the GS degeneracy is 2 or 4
depending to $J_s$. If we compare to the Potts criterion according
to which the transition is of first-order in 2D only when $q > 4$,
then the transition in thin films should be of second order. That
is indeed what we observed.

Another possible cause for the second-order transition observed
here is from the role of the correlation in
 the film. For second-order transitions, some arguments, such as those from
renormalization group, say that the correlation length in the
direction perpendicular  to the film is finite, hence it is
irrelevant to the criticality, the film should have the 2D
character. If a transition is of first order in 3D, i. e. the
correlation length is finite at the transition temperature, then
in thin films the thickness effect may be important: if the
thickness is larger than the correlation length at the transition,
than the first-order transition should remain. On the other hand,
if the thickness is smaller than that correlation length, the
spins then feel an "infinite" correlation length across the film
thickness. As a consequence, two pictures can be thought of: i)
the whole system may be correlated and the first-order character
is to become a second-order one ii) the correlation length is
longer but still finite, the transition remains of first order.

At this point, we would like to emphasize that, in the case of
simple surface conditions, i.e. no significant deviation of the
surface parameters with respect to those of the bulk,  the bulk
behavior is observed when the thickness becomes larger than a few
dozens of atomic layers:\cite{diep79,diep81} surface effects are
insignificant on some thermodynamic properties such as the value
of the critical temperature, the mean value of magnetization at a
given $T$, ... It should be however stressed that the criticality
is very different. It depends on the correlation length compared
to the thickness: for example, we have obtained in the case of
simple cubic films with Ising model the critical exponents
identical to those of  2D Ising universality class up to thickness
of 9 layers.\cite{tu} Due to the small thickness used here, we
think that the 2D character should be assumed.

Now for the anisotropy, remember that in the case studied here, we
do not deal with the discrete Ising model but rather an Ising-like
Heisenberg model. The deviation from the 2D values may then result
in part from a complex coupling between the Ising-like symmetry
and the continuous nature of the classical Heisenberg spins. This
deviation may be important if the anisotropy constant $D$ is small
as in the case  studied here.

To conclude this paragraph, we believe, from physical arguments
given above, that the critical exponents obtained above which do
not belong to any known universality class may result from
different physical mechanisms. This is a subject of future
investigations.






\section{Green's Function Method}

We can rewrite the full Hamiltonian (\ref{eqn:hamil1}) in the
local framework as
\begin{eqnarray}
\mathcal H &=& - \sum_{<i,j>}
J_{i,j}\Bigg\{\frac{1}{4}\left(\cos\theta_{ij} -1\right)
\left(S^+_iS^+_j +S^-_iS^-_j\right)\nonumber\\
&+& \frac{1}{4}\left(\cos\theta_{ij} +1\right) \left(S^+_iS^-_j
+S^-_iS^+_j\right)\nonumber\\
&+&\frac{1}{2}\sin\theta_{ij}\left(S^+_i +S^-_i\right)S^z_j
-\frac{1}{2}\sin\theta_{ij}S^z_i\left(S^+_j
+S^-_j\right)\nonumber\\
&+&\cos\theta_{ij}S^z_iS^z_j\Bigg\}- \sum_{<i>}I_{i}(S^z_i)^2
\label{eq:HGH2}
\end{eqnarray}
where $\cos\left(\theta_{ij}\right)$ is the angle between two NN
spins.

To study properties of quantum spins over a large region of
temperatures, there are only a few methods which give
relatively correct results. Among them, the GF
method  is known to recover the exact results at very low-$T$
obtained from the spin-wave theory. In addition, it is better than the
spin-wave theory at higher temperatures and can be used up to the
transition temperature with of course less precision
on the nature of the phase transition.  It should be emphasized
that the GF method is much better than other methods such as mean-field
theories in estimating the value of the critical temperature. We
choose here this method to study quantum effects at low $T$ and to
obtain the phase diagram at high $T$.

The GF method can be used for non collinear spin
configurations.\cite{Rocco} In the case studied here, one has a
collinear one because of the Ising-like anisotropy. In this case, we define two double-time
GF by\cite{tahir} 
\begin{eqnarray}
G_{ij}(t,t')&=& \ll S^{+}_i(t) ; S^-_j(t')\gg, \\
F_{ij}(t,t')&=&\ll S^{-}_i(t) ; S^+_j(t')\gg.
\end{eqnarray}

The equations of motion for $G_{ij}(t,t')$ and $F_{ij}(t,t')$ are written by
\begin{eqnarray}
i\frac {d}{dt}G_{i,j}\left( t,t'\right) &=& \left<\left[ S^+_i
\left( t\right) , S^-_j \left( t'\right)\right]\right>\delta\left(
t-t'\right) \nonumber\\
&-& \left<\left< \left[\mathcal H, S^+_i\left( t\right)\right] ;
S^-_j \left( t'\right) \right>\right>,
\label{eq:HGEoMG}\\
i\frac {d}{dt}F_{i,j}\left( t,t'\right) &=& \left<\left[ S^-_i
\left( t\right) , S^-_j \left( t'\right)\right]\right>\delta\left(
t-t'\right)\nonumber \\
&-& \left<\left< \left[\mathcal H, S^-_i\left( t\right)\right] ;
S^-_j \left( t'\right) \right>\right>, \label{eq:HGEoMF}
\end{eqnarray}

We shall neglect higher-order correlations by using the Tyablikov
decoupling scheme\cite{Tyablikov} which is known to be valid for
exchange terms.\cite{fro}   Then, we introduce the following Fourier
transforms

\begin{eqnarray}
G_{i, j}\left( t, t'\right) &=& \frac {1}{\Delta}\int\int d\mathbf
k_{xy}\frac{1}{2\pi}\int^{+\infty}_{-\infty}d\omega e^{-i\omega
\left(t-t'\right)}.\nonumber\\
&&\hspace{0.7cm}g_{n,n'}\left(\omega , \mathbf k_{xy}\right)
e^{i\mathbf k_{xy}\cdot \left(\mathbf R_i-\mathbf
R_j\right)},\label{eq:HGFourG}\\
F_{i, j}\left( t, t'\right) &=& \frac {1}{\Delta}\int\int d\mathbf
k_{xy}\frac{1}{2\pi}\int^{+\infty}_{-\infty}d\omega e^{-i\omega
\left(t-t'\right)}.\nonumber\\
&&\hspace{0.7cm}f_{n,n'}\left(\omega , \mathbf k_{xy}\right)
e^{i\mathbf k_{xy}\cdot \left(\mathbf R_i-\mathbf
R_j\right)},\label{eq:HGFourF}
\end{eqnarray}
where $\omega$ is the spin-wave frequency, $\mathbf k_{xy}$
denotes the wave-vector parallel to $xy$ planes, $\mathbf R_i$ is
the position of the spin at the site $i$, $n$ and $n'$ are
respectively the indices of the layers where the sites $i$ and $j$
belong to. The integral over $\mathbf k_{xy}$ is performed in the
first Brillouin zone whose surface is $\Delta$ in the $xy$
reciprocal plane.

The Fourier transforms of the retarded GF satisfy a
set of equations rewritten under the following matrix form
\begin{equation}
\mathbf M \left( \omega \right) \mathbf g = \mathbf u,
\label{eq:HGMatrix}
\end{equation}
where $\mathbf M\left(\omega\right)$ is a square matrix
$\left(2N_z \times 2N_z\right)$, $\mathbf g$ and $\mathbf u$ are
the column matrices which are defined as follows
\begin{equation}
\mathbf g = \left(%
\begin{array}{c}
  g_{1,n'} \\
  f_{1,n'} \\
  \vdots \\
  g_{N_z,n'} \\
  f_{N_z,n'} \\
\end{array}%
\right) , \mathbf u =\left(%
\begin{array}{c}
  2 \left< S^z_1\right>\delta_{1,n'} \\
  0 \\
  \vdots \\
  2 \left< S^z_{N_z}\right>\delta_{N_z,n'} \\
  0 \\
\end{array}%
\right) , \label{eq:HGMatrixgu}
\end{equation}
and
\begin{equation}
\mathbf M\left(\omega\right) = \left(%
\begin{array}{ccccc}
  A^+_1    & B_1    &  D^+_1 &  D^-_1 & \cdots \\
  -B_1     & A^-_1  & -D^-_1 & -D^+_1 & \vdots \\
   \vdots  & \cdots & \cdots & \cdots &\vdots\\
  \vdots   & C^+_{N_z}   & C^-_{N_z}   & A^+_{N_z}      & B_{N_z}\\
  \cdots        & -C^-_{N_z}  & -C^+_{N_z}  & -B_{N_z}       & A^-_{N_z}\\
\end{array}%
\right), \label{eq:HGMatrixM}
\end{equation}
where
\begin{eqnarray}
A_n^\pm &=&  \omega \pm\Big[\frac{1}{2}J_n \left< S^z_n\right>
\left(Z\gamma\right)\left(\cos\theta_{n} +1\right)\nonumber\\
&-& J_n \left< S^z_n\right>Z\cos\theta_{n} -2I_{n} \left< S^z_n\right>\nonumber\\
&-&2 J_{n, n+1}\left< S^z_{n+1}\right>\cos\theta_{n,n+1}^{(a)} \nonumber\\
&-&2 J_{n, n+1}\left< S^z_{n+1}\right>\cos\theta_{n,n+1}^{(b)} \nonumber\\
&-&2 J_{n, n-1}\left< S^z_{n-1}\right>\cos\theta_{n,n-1}^{(a)}\nonumber\\
&-&2 J_{n, n-1}\left< S^z_{n-1}\right>\cos\theta_{n,n-1}^{(b)} \Big],\\
B_n &=& \frac{1}{2}J_{n}\left<
S^z_n\right>\left(Z\gamma\right)\left(\cos\theta_{n}-1\right),\\
C_n^\pm &=& J_{n,n-1}\left<
S^z_n\right>\left(\cos\theta_{n,n-1}^{(a)}\pm 1\right)\nonumber\\
&+& J_{n,n-1}\left<
S^z_n\right>\left(\cos\theta_{n,n-1}^{(b)}\pm 1\right),\\
D_n^\pm &=& J_{n,n+1}\left<
S^z_n\right>\left(\cos\theta_{n,n+1}^{(a)}\pm 1\right)\nonumber\\
&+&J_{n,n+1}\left<S^z_n\right>\left(\cos\theta_{n,n+1}^{(b)}\pm
1\right),
\end{eqnarray}
in which, $Z=4$ is the number of in-plane NN, $\theta_{n,n\pm
1}^{(a)}$ the angle between two NN spins of sublattice 1 and 3
belonging to the layers $n$ and $n\pm1$ (see Fig.
\ref{fig:gsstruct}), $\theta_{n,n\pm 1}^{(b)}$ the angle between two NN
spins of sublattice 1 and 4, $\theta_{n}$ the angle between
two in-plane NN  spins in the layer $n$, and
$$\gamma =\frac{1}{Z}\left[ 4\cos \left( \frac{k_x a}{2} \right)\cos \left( \frac{k_y a}{2} \right)\right].$$

Here, for compactness we have used the following notations:

i) $J_n$ and $D_n$ are the in-plane interactions. In the present
model $J_n$ is equal to $J_s$ for the two surface layers and equal
to $J$ for the interior layers. All $D_n$ are set to be $D$.

ii) $ J_{n,n\pm 1}$ are the interactions between a spin in the
$n$-th layer and its neighbor in the $(n\pm 1)$-th layer. Of
course, $ J_{n,n-1}=0$ if $n=1$, $ J_{n,n+1}=0$ if $n=N_z$.

Solving det$|\mathbf M|=0$, we obtain the spin-wave spectrum
$\omega$ of the present system.  The solution for the GF $g_{n,n}$ is given by
\begin{equation}
g_{n,n} = \frac{\left|\mathbf M\right|_n}{\left|\mathbf M\right|},
\end{equation}
with $\left|\mathbf M\right|_n$ is the determinant made by
replacing the $n$-th column of $\left|\mathbf M\right|$ by
$\mathbf u$ in (\ref{eq:HGMatrixgu}). Writing now
\begin{equation}
\left|\mathbf M\right| = \prod_i \left(\omega -
\omega_i\left(\mathbf k_{xy}\right)\right),
\end{equation}
one sees that $\omega_i\left(\mathbf k_{xy}\right) ,\ i = 1,\cdots
,\ N_z$, are poles of the GF $g_{n,n}$.
$\omega_i\left(\mathbf k_{xy}\right)$ can be obtained by solving
$\left|\mathbf M\right|=0$. In this case, $g_{n,n}$ can be
expressed as
\begin{equation}
g_{n, n} = \sum_i\frac {f_n\left(\omega_i\left(\mathbf
k_{xy}\right)\right)}{\left( \omega - \omega_i\left(\mathbf
k_{xy}\right)\right)}, \label{eq:HGGnn}
\end{equation}
where $f_n\left(\omega_i\left(\mathbf k_{xy}\right)\right)$ is
\begin{equation}
f_n\left(\omega_i\left(\mathbf k_{xy}\right)\right) = \frac{\left|
\mathbf M\right|_n \left(\omega_i\left(\mathbf
k_{xy}\right)\right)}{\prod_{j\neq i}\left(\omega_j\left(\mathbf
k_{xy}\right)-\omega_i\left(\mathbf k_{xy}\right)\right)}.
\end{equation}

Next, using the spectral theorem which relates the correlation
function \(\langle S^-_i S^+_j\rangle \) to the GF,\cite{zu} one has
\begin{eqnarray}
\left< S^-_i S^+_j\right> &=& \lim_{\varepsilon\rightarrow 0}
\frac{1}{\Delta}\int\int d\mathbf k_{xy}
\int^{+\infty}_{-\infty}\frac{i}{2\pi}\big( g_{n, n'}\left(\omega
+ i\varepsilon\right)\nonumber\\
&-& g_{n, n'}\left(\omega - i\varepsilon\right)\big)
\cdot\frac{d\omega}{e^{\beta\omega} - 1}e^{i\mathbf
k_{xy}\cdot\left(\mathbf R_i -\mathbf R_j\right)},
\end{eqnarray}
where $\epsilon$ is an  infinitesimal positive constant and
$\beta=1/k_BT$, $k_B$ being the Boltzmann constant.

Using the GF presented above, we can calculate
self-consistently various physical quantities as functions of
temperature $T$.   We  start the self-consistent calculation from
$T=0$ with a small step for temperature: $5\times 10^{-3}$ at low $T$ and
$10^{-1}$ near $T_c$ (in units of $J/k_B$). The convergence precision has
been fixed at the fourth figure of the values obtained for the
layer magnetizations. We know from the previous section that the
spin configuration is collinear, therefore in this section, we
shall use a large value of Ising anisotropy $D$ in order to get
a rapid numerical convergence. For numerical calculation,
we will use $D=4$ and $J=-1$ and a size of $80^2$ points in the
first Brillouin zone.

\begin{figure}[htb!]
\centerline{\epsfig{file=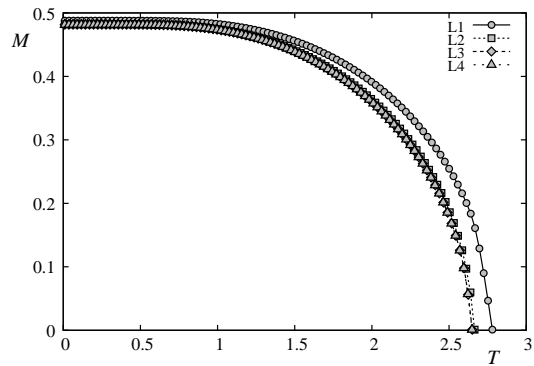,width=2.8in}} \caption{Layer
magnetization of first four layers vs temperature for $J_s = -1.0$
and $D = 4$. $L_j$ denotes the sublattice magnetization of layer
$j$. Note that except the first layer, all other layer magnetizations 
coincide in this figure scale.} \label{fig:GFJ10}
\end{figure}

\begin{figure}[htb!]
\centerline{\epsfig{file=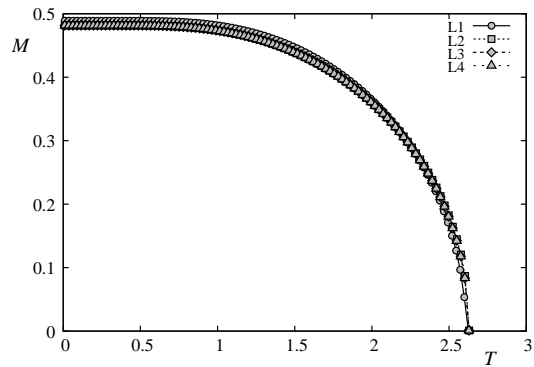,width=2.8in}} \caption{Layer
magnetizations of first four layers vs temperature for $J_s = -0.8$
and $D = 4$. $L_j$ denotes the sublattice magnetization of layer
$j$.  Note that all layer magnetizations coincide in this figure scale.} \label{fig:GFJ08}
\end{figure}

Figure  \ref{fig:GFJ10} shows the sublattice magnetizations of the
first four layers. As seen, the first-layer one is larger than the
other three just as in the case of the classical spins shown in
Fig. \ref{fig:N24D01J10}.   This difference in sublattice
magnetization between layers vanishes at $J_s \simeq -0.8$ as seen
in Fig. \ref{fig:GFJ08}. Again here, one has a good agreement with
the case of classical spins shown in Fig.  \ref{fig:N24D01J08}.

\begin{figure}[htb!]
\centerline{\epsfig{file=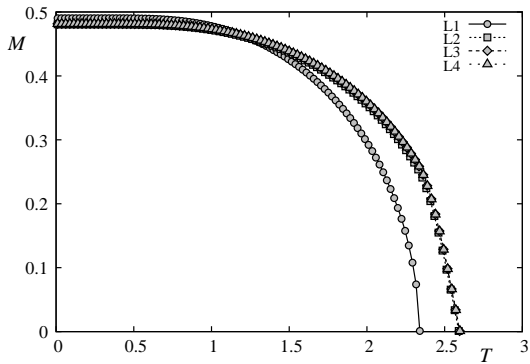,width=2.8in}} \caption{Layer
magnetization of first four layers vs temperature for $J_s = -0.5$
and $D = 4$. $L_j$ denotes the sublattice magnetization of layer
$j$. Note that except the first layer, all other layer magnetizations 
coincide in this figure scale. See text for comments on the crossover of surface
magnetization.} \label{fig:GFJ05}
\end{figure}

For $J_s>-0.8$,  the sublattice magnetization of the first layer
is larger  at low $T$ and higher at high $T$ as seen in Fig.
\ref{fig:GFJ05} for $J_s=-0.5$. This crossover of sublattice
magnetizations comes from the competition between quantum
fluctuations and the strength of $J_s$: when $|J_s|$ is small,
quantum fluctuations of the surface layer are small yielding a
small zero-point spin contraction for surface spins at $T=0$. So, surface 
magnetization is higher than the interior ones.  At
higher $T$, however, small $|J_s|$ gives rise to a small local field for
surface spins which in turn yields a smaller surface magnetization at high
$T$.  This crossover has been found earlier in antiferromagnetic
superlattices and films.\cite{Diep1989sl,diep91-af-films}

\begin{figure}[htb!]
\centerline{\epsfig{file=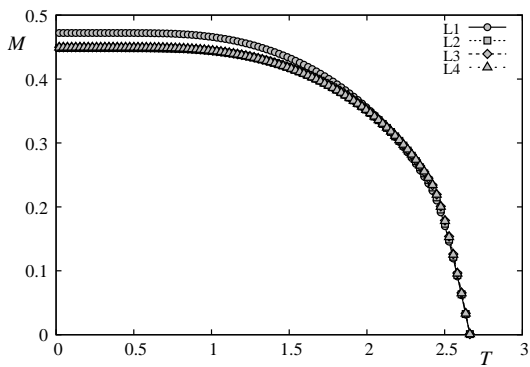,width=2.8in}} \caption{Layer
magnetization of first four layers vs temperature for $J_s = -0.1$
and $D = 4$. $L_j$ denotes the sublattice magnetization of layer
$j$. Only at low $T$ the surface magnetization is distinct from 
the other ones.} \label{fig:GFJ01}
\end{figure}

For $J_s=-0.1$, there is no more crossover at low $T$ as seen in Fig. \ref{fig:GFJ01}.  
Moreover, there is only a single transition at $T_c\simeq 2.65$ for both surface and interior layers.

\begin{figure}[htb!]
\centerline{\epsfig{file=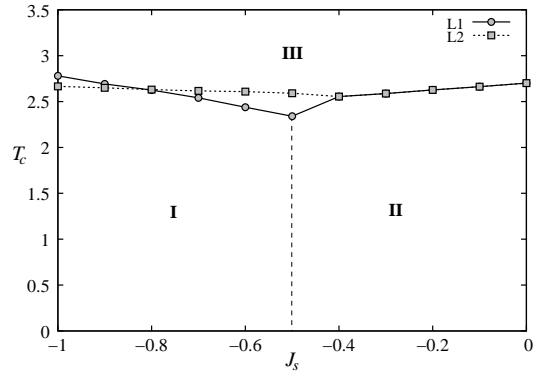,width=2.8in}} \caption{Phase
diagram obtained by the Green's function method with $D = 4$. 
$L_j$ denotes the transition temperature of the sublattice magnetization
of layer $j$. See text for comments.} \label{fig:GFPD}
\end{figure}

We summarize in Fig. \ref{fig:GFPD} the phase diagram for the quantum spin 
case obtained with the GF method.  The vertical discontinued line indicates the boundary
between ordered phases of types I and II.  Phase III is paramagnetic. Note the 
following interesting points:

i) for
$J_s<-0.4$ there is a surface transition distinct from that of interior layers,

ii) for $J_s<-0.8$, 
surface transition occurs at a temperature higher than that of interior layers,

iii) there is a reentrance between $J_s=-0.4$ and $J_s=-0.5$. This is very similar to  
the phase diagram of the classical spins obtained by MC simulations shown in Fig.  \ref{fig:PDJ24D01}.

\section{Concluding Remarks}

We have studied in this paper the properties of a thin film made from 
a fully frustrated material, namely the FCC antiferromagnet.  We have considered both classical and 
quantum Heisenberg spin model with an Ising-like single-ion anisotropy.  The classical case was
treated by Monte Carlo simulation while the quantum case was studied by
the Green's function method.
Several important results are found in this paper. We found that 
the presence of a surface reduces the GS degeneracy of the fully-frustrated FCC
antiferromagnet and  there exists a critical value of the in-plane surface interaction
$J_s^c=-0.5$ which separates the GS configuration of type I from that of
type II. We have studied the phase transition of the system. 
The surface spin ordering is destroyed in general at a
temperature different from that of the interior layers. We found that
in a small region just above $J_s^c$ there is a reentrant
phase: with decreasing $T$ the system first changes from the paramagnetic phase to
the type II phase, and then enters at a lower temperature into the type I phase.
The critical behaviors of surface and interior layers have been shown and discussed.

We hope that these unusual surface properties will help experimentalists to
analyze their data obtained for real systems where frustration plays an important role. 

One of us (VTN) thanks the "Asia Pacific Center for Theoretical Physics" (South Korea) 
for a financial
post-doc support and hospitality during the period 2005-2006 where
part of this work was carried out.

{}

\end{document}